# Metal-Enhanced Fluorescence of Carbon Nanotubes

Guosong Hong, Scott M. Tabakman, Kevin Welsher, Hailiang Wang, Xinran Wang, and Hongjie Dai*

*Department of Chemistry, Stanford University, Stanford, California 94305*

***Abstract:*** The photoluminescence (PL) quantum yield of single-walled carbon nanotubes (SWNTs) is relatively low, with various quenching effects by metallic species reported in the literature. Here, we report the first case of metal enhanced fluorescence (MEF) of surfactant-coated carbon nanotubes on nanostructured gold substrates. The photoluminescence quantum yield of SWNTs is observed to be enhanced more than 10-fold. The dependence of fluorescence enhancement on metal-nanotube distance and on surface plasmon resonance (SPR) of the gold substrate for various SWNT chiralities is measured to reveal the mechanism of enhancement. Surfactant-coated SWNTs in direct contact with metal exhibit strong MEF without quenching, suggesting a small quenching distance for SWNTs on the order of the van der Waals distance, beyond which the intrinsically fast non-radiative decay rate in nanotubes is little enhanced by metal. The metal enhanced fluorescence of SWNTs is attributed to radiative lifetime shortening through resonance coupling of SWNT emission to the re-radiating dipolar plasmonic modes in the metal.

Semiconducting single-walled carbon nanotubes (SWNTs) with various (m,n) indices, or chiralities, are quasi one-dimensional materials exhibiting characteristic optical absorption and photoluminescence (PL).[1] The intrinsic photoluminescence of SWNTs in the near infrared (NIR) above 1 μm makes them promising candidates for biological imaging *in vitro*[2] and *in vivo*.[3] The fluorescence quantum yields of SWNTs are relatively low compared to organic fluorophores or quantum dots emitting below 1 μm due to the nonradiative nature of the lowest-energy excitons,[4] and extrinsic factors that tend to quench the nanotube fluorescence. Metallic SWNTs in nanotube bundles, metal substrates and $SiO_2$ substrates in close proximity to SWNTs and adsorbed molecular species in acidic pH have all been reported to quench SWNT fluorescence through energy transfer or charge transfer mechanisms.[5]

On the other hand, metal-enhanced fluorescence (MEF), a phenomenon reported for various organic fluorophores[6] and quantum dots[7] proximal to gold or silver substrates or other nanostructures, has not been observed with carbon nanotubes thus far. Here, we report the first case of MEF of surfactant coated, water soluble SWNTs on solution-grown Au films seeded by Au nanoparticles (called ‘AuAu films’).[8] We find that surfactant coated SWNTs in contact with a metal substrate are immune to quenching, likely due to protection by the coating layer. To the contrary, we observed enhanced fluorescence of SWNTs by more than 10-fold. The enhancement factor monotonically decreases when the SWNTs are placed further away from the AuAu surface using ‘spacing’ layers of self-assembled alkanethiol monolayers or thin $Al_2O_3$ layers grown on Au by atomic layer deposition (ALD). Also, we find that the degree of fluorescence enhancement for various chirality SWNTs increases monotonically with the AuAu substrate plasmonic absorbance/extinction at the emission wavelengths of the nanotubes. The results suggest that the mechanism of SWNT fluorescence enhancement is due to radiative lifetime shortening of the excited state, resulting from resonant coupling of nanotube emission with the scattering and re-radiating component of plasmons localized on the surface of the metal substrate.

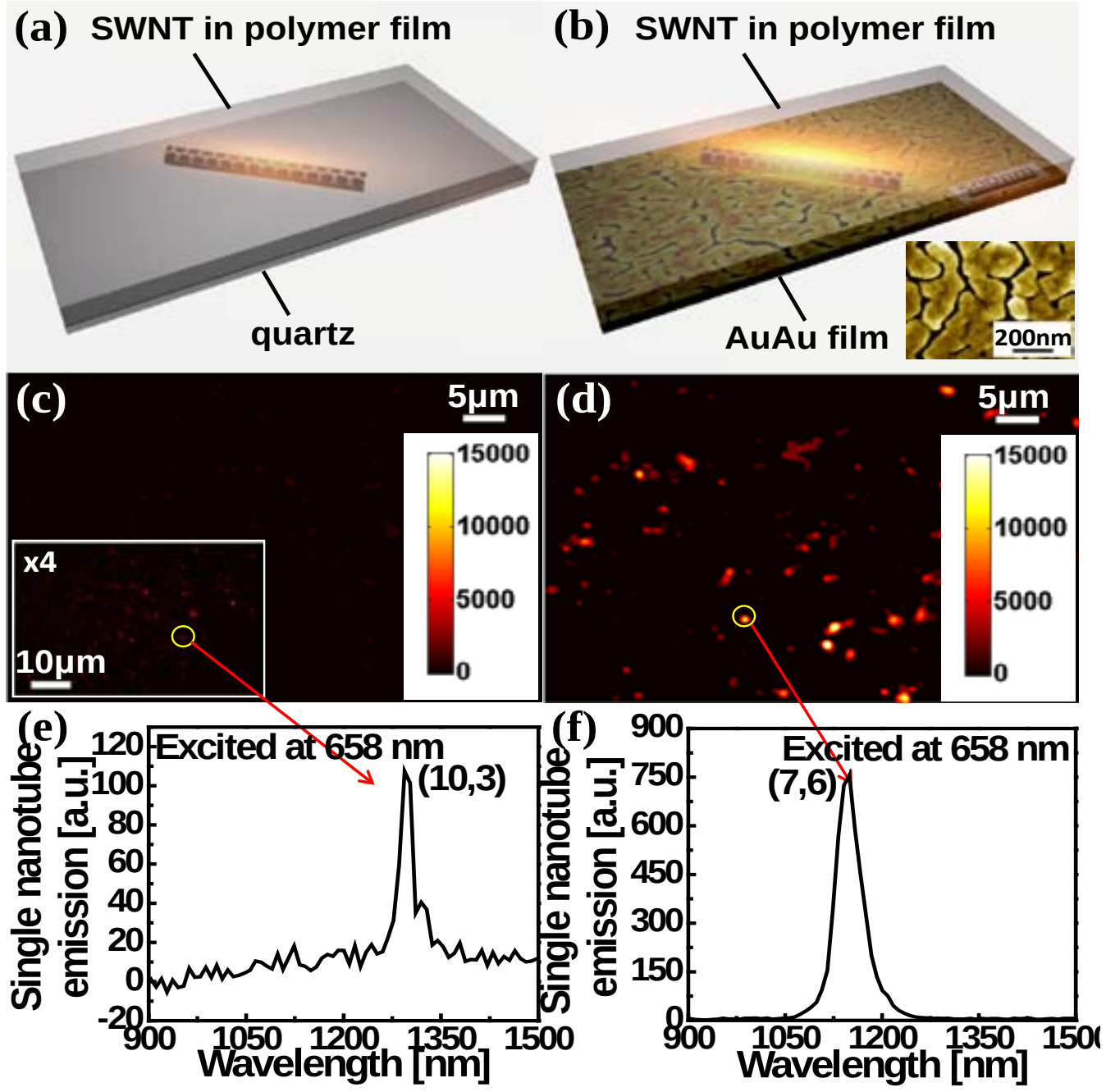

***Figure 1.*** Schematics showing SWNTs embedded in a PVP matrix on (a) quartz and (b) an AuAu film made on glass. The inset of (b) shows an SEM image of the nanostructured AuAu film. (c) NIR photoluminescence image of SWNTs on quartz and (d) on AuAu film at the same emission intensity scale showing much enhanced fluorescence on AuAu. The inset in (c) shows the fluorescence image of (c) after amplifying the intensity by 4 times. (e)&(f) Emission spectra (at 658 nm excitation) for two individual nanotubes in (c) on quartz and in (d) on AuAu substrate respectively. Note that the two different chirality (10,3) and (7,6) tube both have absorption band near the 658 nm excitation.

SWNTs (average length ~400 nm)[3] coated by PEGylated phospholipid (DSPE-mPEG)[9] in a stable water suspension were prepared by exchanging sodium cholate solubilized SWNTs into DSPE-mPEG coating to afford relatively long SWNTs with optimal quantum yield.[3] The SWNT suspension was mixed with polyvinylpyrrolidone (PVP) and spin-coated onto both a quartz substrate (Fig.1a) and an AuAu substrate (Fig.1b) made by solution-phase seeding and growth of gold nanoparticles,[8] as described in the supporting information (SI). The morphology of as-made AuAu substrates imaged by scanning electron microscopy (SEM, Fig.1b inset) showed small gold islands with small gaps in between the islands. The substrates have been shown recently to afford excellent surface-enhanced Raman scattering (SERS) properties for nanotubes and other molecules.[8] Using a 658 nm laser excitation and an InGaAs 2D camera, we obtained fluorescence images of ensembles of SWNTs in the 0.9-1.7 μm emission range and detected significantly brighter fluorescence on the AuAu substrate (Fig.1d) than on bare quartz (Fig.1c) for the same density of SWNTs deposited on both substrates. By taking emission spectra for ensembles of SWNTs in the same field of view on both substrates (Fig.S1), we found that this particular AuAu film enhanced the overall fluorescence intensity (integrated in the 0.9-1.7 μm range) of SWNTs by ~8 times compared to SWNTs on bare quartz.

We identified and characterized individual SWNTs on both quartz (Fig.1e) and AuAu (Fig.1f) substrates. Under a laser excitation with rotating polarization, individual nanotubes exhibited fluorescence emission periodic to the polarization angle (periodicity~180°, Fig.S2), as shown previously[10] with maximum emission occurring when the laser polarization was parallel to the nanotube axis. We recorded emission spectra for individual SWNTs and assigned (m,n) indices (Fig.1e,1f) based on established spectral assignments[1b]. Due to the significantly enhanced fluorescence, individual SWNTs on the AuAu substrate were easily identified and characterized, with emission spectra showing much higher signal/noise ratios, smoother profiles and less noisy baselines. The comparisons clearly suggested the advantages of using the AuAu substrate for the characterization of relatively weak fluorophores.

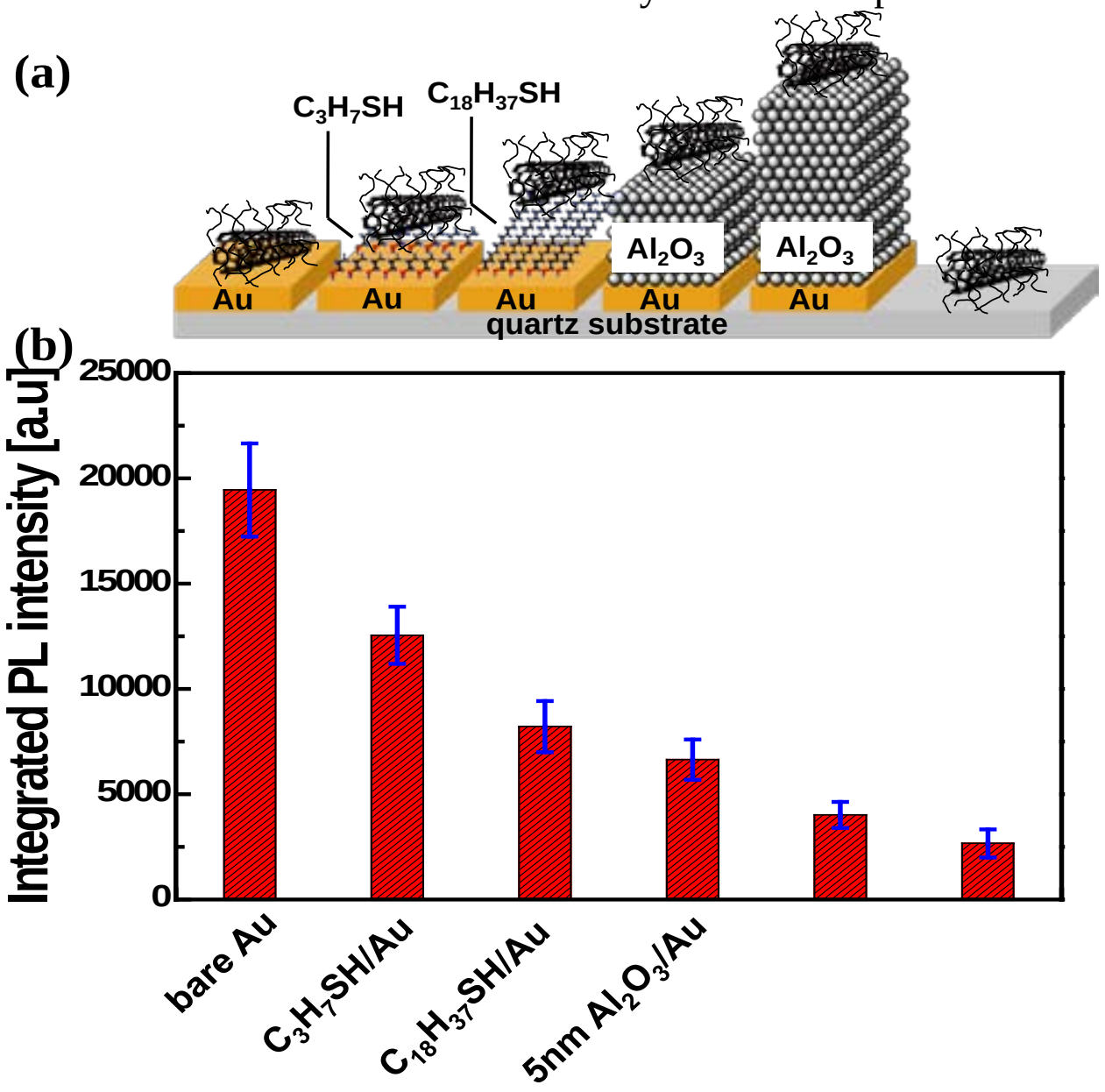

***Figure 2.*** (a) A schematic drawing showing SWNTs on various substrates indicated. (b) A bar chart showing the decreasing PL intensity of SWNTs on different substrates, corresponding to those in the upper schematic.

To elucidate the mechanism of the observed fluorescence enhancement of SWNTs, we formed various spacer layers on AuAu substrates and measured the degree of SWNT fluorescence enhancement as a function of the thickness of the spacers including self-assembled monolayers (SAMs) of 1-propanethiol (thickness~0.5 nm) and 1-octadecanethiol (~2 nm), and ~5 nm $Al_2O_3$ and ~10 nm $Al_2O_3$ made by ALD on cysteamine coated AuAu substrates (Figure 2a). For the same amount of SWNTs deposited onto these substrates, we observed monotonic decrease of SWNT fluorescence (Figure S3) as the spacer layer thickness increased, with SWNTs on bare quartz exhibiting the weakest fluorescence (Fig.2b). This result suggested that the nanostructured Au film was responsible for enhanced fluorescence of SWNTs. The enhancement effect reduced as SWNTs were spaced away from the underlying metal surface.

We attribute the enhanced fluorescence of SWNTs on our AuAu substrates to resonance coupling of nanotube emission with surface plasmons of the underlying AuAu film. Similar to MEF of common fluorophores, coupling between nanotube fluorescence emission and the plasmon modes in metal structures shortens the radiative lifetime ($1/\Gamma_r$) of excited states, leading to higher quantum yield η [$\eta=\Gamma_r/(\Gamma_r+\Gamma_{nr})$, where $1/\Gamma_{nr}$=non-radiative lifetime].[11,12] Such coupling is well known to reduce monotonically as fluorophores are placed away from the metal substrates by spacer layers,[12] which reduces the enhancement effect as seen in our experiment. SWNTs are known to exhibit low intrinsic quantum efficiency on the order of ~1-3% in the absence of any metal enhancement, corresponding to $\Gamma_r \ll \Gamma_{nr}$ and $\eta\approx\Gamma_r/\Gamma_{nr}$. If the intrinsically high non-radiative decay rate $\Gamma_{nr}$ is not further enhanced by metal, then any enhanced radiative decay $\Gamma_r$ will translate into higher quantum efficiencies of SWNTs, by up to ~8X as in Fig.1 and 2.

It is known that coupling between fluorophores and the non-radiating dark plasmon modes (non-dipolar higher order modes[12b]) in metal could enhance non-radiative decay rate and shorten the non-radiative lifetimes ($1/\Gamma_{nr}$) due to energy transfer, leading to reduced fluorescence emission or even quenching.[12,13] In our experiments, no quenching was observed for the most proximal case of SWNTs on AuAu substrates without any intentionally placed spacers. However, even in this case there was a ~3 nm distance between SWNTs and gold corresponding to the radius of gyration of the DSPE-mPEG (molecular weight~5000 Da) polymer chains coated on SWNTs.[14] This prevented SWNTs from direct contact (within the van der Waals distance of ~0.34 nm of graphitic materials) with gold and thus avoided quenching. We have also coated SWNTs with a much smaller surfactant molecule cholate (molecular weight ~400 Da) and also observed MEF of these tubes on AuAu substrate vs. on quartz. These results suggest that the 'quenching distance' on metal surfaces for SWNTs seems to be shorter than that of most organic fluorophores (7-15nm[15]). The intrinsically high non-radiative decay rate $\Gamma_{nr}$ of SWNT excited states appears to be only significantly enhanced when nanotubes are very close to metallic species, like in a bundle where metallic tubes intimately coexist. We conclude that SWNTs with non-covalent molecular functionalization are in fact relatively immune to quenching with a small quenching distance of << ~3 nm to metal surfaces.

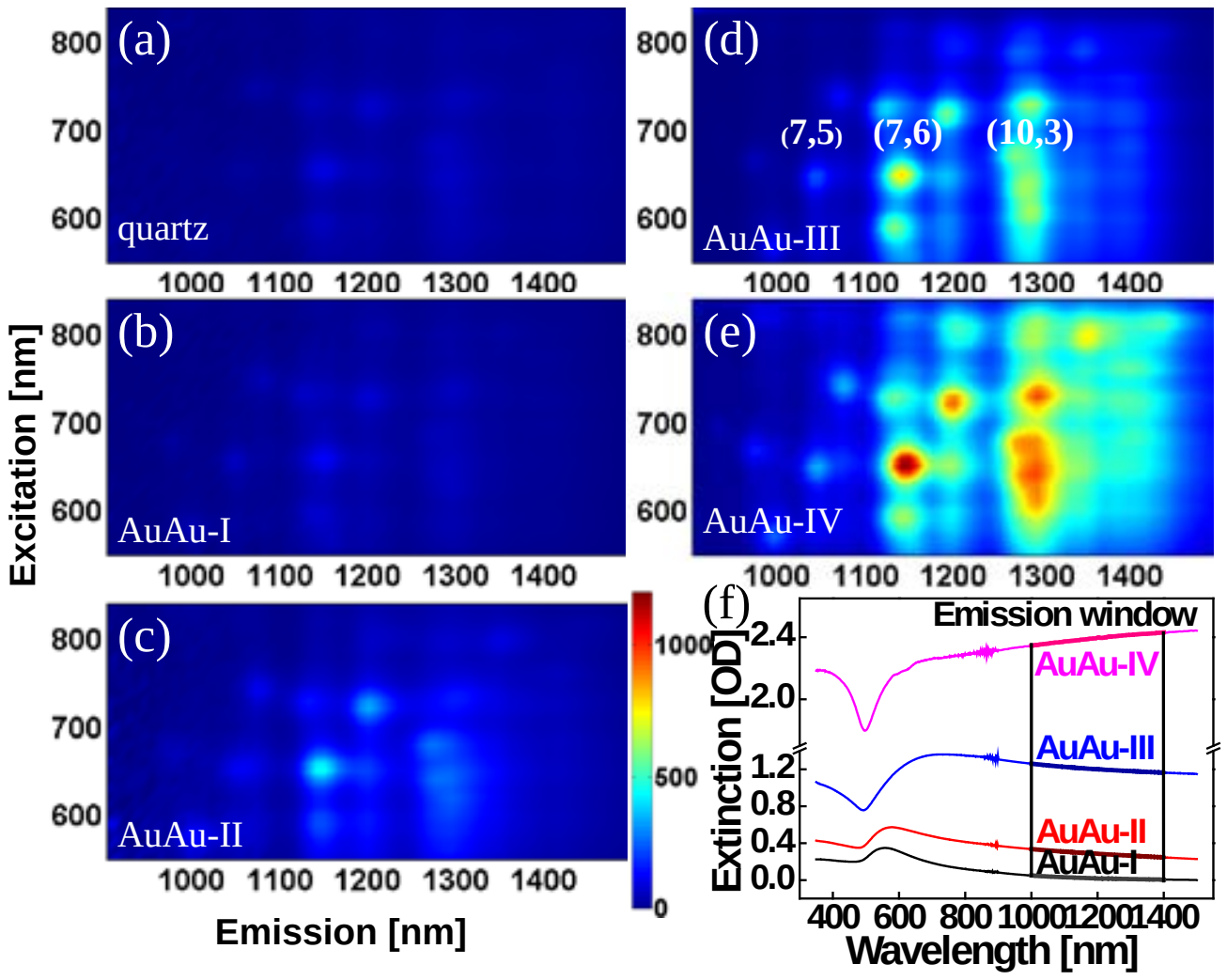

***Figure 3.*** Photoluminescence excitation (PLE) spectra of SWNTs on (a) quartz, (b) AuAu-I, (c) AuAu-II, (d) AuAu-III and (e) AuAu-IV substrates with increasing Au coverage and thickness. These AuAu substrates are made at different conditions, according to the experimental details in SI, so that they have increasing extinction in the UV-Vis-NIR spectra in (f).

We investigated MEF of SWNTs on four different AuAu substrates (labeled as AuAu-I to -IV, Fig.3, see SI for their synthesis) with increasing metal coverage or thickness on glass substrates, with higher optical absorbance and increasingly red-shifted surface plasmon resonance features (Fig.3f). Photoluminescence excitation (PLE) spectra of the same amount of SWNTs deposited on bare quartz (Fig.3a) and the four AuAu films (Fig.3b-e) revealed a monotonic increase in PL intensity for various chirality SWNTs (Fig.3&4), and the enhancement factor was up to 10-12X on the thickest AuAu film (Fig.3e&4d). Noteworthy was that a similar trend was observed for MEF of indocyanine green (ICG) on silver substrates, where higher fluorescence was detected on thicker silver films with higher optical absorbance.[16]

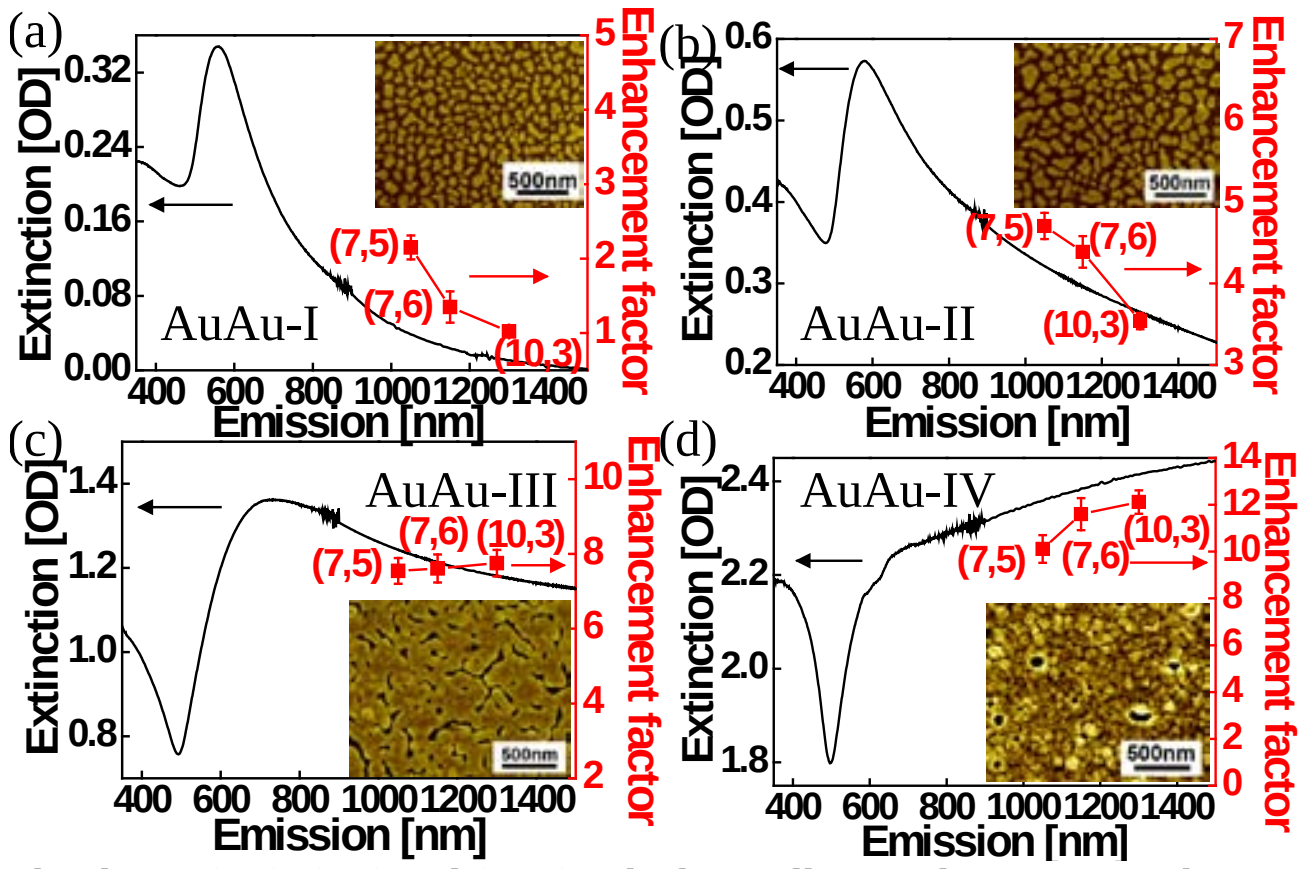

***Figure 4.*** Enhancement factors of three chiralities, (7,5), (7,6) and (10,3), which are all excited near 650 nm but emitting in range from 1050 nm to 1300 nm, plotted against their emissions, alongside with the UV-Vis-NIR extinction curves for (a) AuAu-I, (b) AuAu-II, (c) AuAu-III and (d) AuAu-IV substrates. Enhancement factors are shown in red squares, while UV-Vis-NIR curves are shown in black. The error bars for EF of each chirality were obtained by taking the standard deviation of the brightest and eight surrounding pixels within one spot of a corresponding chirality in the PLE plots.

We analyzed MEF of several groups of (m,n) SWNTs with similar excitation wavelength but different emissions [e.g. (7,5), (7,6) and (10,3) tubes, Fig.3d], and correlated their MEF enhancement factors with the SPR absorbance at the emission wavelengths. Figure 4a-c show decreasing surface plasmons at longer wavelengths for the first three AuAu substrates, whereas AuAu-IV shows an increasing trend. The MEF enhancement factors of (7,5), (7,6) and (10,3) tubes exhibited similar trends as the SPR absorbance, decreasing with emission wavelength on AuAu-I and AuAu-II and increasing with emission wavelength on AuAu-IV (Fig.4a,b,d). In particular, the thinnest AuAu-I substrate showed SPR absorbance approaching zero at the emission wavelength of certain tubes [e.g., (10,3) tube], indicating no surface plasmon resonance on the AuAu-I film in this wavelength regime. In this case, fluorescence was barely enhanced for the (10,3) tube with an EF = ~1, even though the AuAu-I film exhibited substantial optical absorbance at the excitation wavelength of the (10,3) tube. These results suggested that MEF of SWNTs was due to resonance coupling of SWNT fluorescence emission (rather than excitation) with surface plasmons in the underlying gold film.

The (7,5), (7,6) and (10,3) tubes on AuAu-III film showed similar enhancement factors but appeared to be slightly against the trend of the SPR absorbance/extinction curve (Fig.4c). This was attributed to the small variations in the extinction in the 1000-1400 nm range and that the intrinsic chirality dependence of MEF could be at play. We also analyzed SWNTs with similar emission wavelengths and different excitation wavelengths. The plotted EF's vs. excitation wavelengths, alongside with SPR extinction curves for all AuAu substrates, show much worse correlation (Figure S4) than that in Fig.4.

It is established that fluorescence enhancement of a molecule by metal depends on the size (and plasmonic frequency) of the metal structure. Stronger dipolar coupling between the molecule and plasmons in the metal prefers smaller metal sizes.[12] On the other hand, larger metal structures, such as in the thicker AuAu films, are submicron structures composed of smaller nanocrystals,[17] and they afford a stronger scattering or re-radiating component of the optical extinction, leading to higher enhancement by re-radiating the emission.[12b] These two competing factors often lead to an optimum metal size for MEF of molecules. In our case with nanotubes of ~400nm length as studied here, the monotonic increase in MEF as the AuAu film thickens suggests that stronger scattering and re-radiating effects of surface plasmons are dominant and responsible for the enhanced SWNT fluorescence. We also studied the MEF of shorter nanotubes (~ 150nm on average) with much lower quantum yield (made by direct sonication in DSPE-mPEG surfactant),[3] and observed even higher fluorescence enhancement factors (up to 40X) on relatively thin AuAu-III film with many small gaps between the Au islands. On thicker and more continuous AuAu films, a decrease in enhancement factor was observed for these short tubes. These results suggested short SWNTs near the gaps of Au nano-islands may exhibit fluorescence enhancement due to electric field enhancement in addition to resonance plasmonic enhancement.

In summary, we observed the first metal enhanced fluorescence of SWNTs resulted from radiative lifetime shortening through resonance coupling of SWNT emission to plasmonic modes in the metal. The enhancement effect decreases monotonically as the separation between nanotubes and metal increases. Surfactant-coated SWNTs in direct contact with metal exhibit the strongest MEF without quenching, suggesting a small quenching distance on the order of the van der Waals distance, beyond which the intrinsically fast non-radiative decay rate in nanotubes is little changed. SWNT fluorescence enhancement monotonically increases with optical extinction of thicker gold films due to increased scattering and re-radiating of stronger surface plasmons resonantly coupled to nanotube emission. We believe that metal enhanced fluorescence of SWNTs by > 10 times will be welcoming news for these materials in fundamental and practical applications including sensing, detection and imaging.

**Acknowledgement.** This work was supported by NIH-NCI 5R01CA135109-02, CCNE-TR and NSF CHE-0639053.

## References

(1) (a) O'Connell, M. J.; Bachilo, S. M.; Huffman, C. B.; Moore, V. C.; Strano, M. S.; Haroz, E. H.; Rialon, K. L.; Boul, P. J.; Noon, W. H.; Kittrell, C.; Ma, J.; Hauge, R. H.; Weisman, R. B.; Smalley, R. E. *Science* **2002,** *297,* 593. (b) Bachilo, S. M.; Strano, M. S.; Kittrell, C.; Hauge, R. H.; Smalley, R. E.; Weisman, R. B. *Science* **2002,** *298,* 2361.

(2) (a) Cherukuri, P.; Bachilo, S. M.; Litovsky, S. H.; Weisman, R. B. *J. Am. Chem. Soc.* **2004,** *126,* 15638. (b) Welsher, K.; Liu, Z.; Daranciang, D.; Dai, H. *Nano Lett.* **2008,** *8,* 586. (c) Liu, Z.; Tabakman, S.; Welsher, K.; Dai, H. *Nano Res.* **2009,** *2,* 85.

(3) Welsher, K.; Liu, Z.; Sherlock, S. P.; Robinson, J. T.; Chen, Z.; Daranciang, D.; Dai, H. *Nat. Nanotech.* **2009,** *4,* 773.

(4) Perebeinos, V.; Tersoff, J.; Avouris, P. *Nano Lett.* **2005,** *5,* 2495.

(5) (a) Lauret, J. S.; Voisin, C.; Cassabois, G.; Delalande, C.; Roussignol, P.; Jost, O.; Capes, L. *Phys. Rev. Lett.* **2003,** *90,* 057404. (b) Lefebvre, J.; Homma, Y.; Finnie, P. *Phys. Rev. Lett.* **2003,** *90,* 217401. (c) Lefebvre, J.; Austing, D. G.; Bond, J.; Finnie, P. *Nano Lett.* **2006,** *6,* 1603. (d) Cognet, L.; Tsyboulski, D. A.; Rocha, J. R.; Doyle, C. D.; Tour, J. M.; Weisman, R. B. *Science* **2007,** *316,* 1465.

(6) Sokolov, K.; Chumanov, G.; Cotton, T. M. *Anal. Chem.* **1998,** *70,* 3898.

(7) (a) Shimizu, K. T.; Woo, W. K.; Fisher, B. R.; Eisler, H. J.; Bawendi, M. G. *Phys. Rev. Lett.* **2002,** *89,* 117401. (b) Ray, K.; Badugu, R.; Lakowicz, J. R. *J. Am. Chem. Soc.* **2006,** *128,* 8998.

(8) Tabakman, S. M.; Chen, Z.; Casalongue, H. S.; Wang, H.; Dai, H. submitted.

(9) (a) Liu, Z.; Cai, W.; He, L.; Nakayama, N.; Chen, K.; Sun, X.; Chen, X.; Dai, H. *Nat. Nanotech.* **2007,** *2,* 47. (b) Liu, Z.; Tabakman, S. M.; Chen, Z.; Dai, H. *Nat. Protoc.* **2009,** *4,* 1372. (c) Liu, Z.; Tabakman, S.; Sherlock, S.; Li, X.; Chen, Z.; Jiang, K.; Fan, S.; Dai, H. *Nano Res.* **2010,** *3,* 222.

(10) Tsyboulsky, D. A.; Bachilo, S. M.; Weisman, R. B. *Nano Lett.* **2005,** *5,* 975.

(11) Geddes, C. D.; Lakowicz, J. R. *J. Fluoresc.* **2002,** *12,* 121.

(12) (a) Gersten, J.; Nitzan, A. *J. Chem. Phys.* **1981,** *75,* 1139. (b) Mertens, H.; Koenderink, A. F.; Polman, A. *Phys. Rev. B* **2007,** *76,* 115123.

(13) Dulkeith, E.; Morteani, A. C.; Niedereichholz, T.; Klar, T. A.; Feldmann, J. *Phys. Rev. Lett.* **2002,** *89,* 203002.

(14) Bhat, R.; Timasheff, S. N. *Protein Sci.* **1992,** *1,* 1133.

(15) Chhabra, R.; Sharma, J.; Wang, H.; Zou, S.; Lin, S.; Yan, H.; Lindsay, S.; Liu, Y. *Nanotech.,* **2009,** *20,* 485201.

(16) Aslan, K.; Leonenko, Z.; Lacowicz, J. R.; Geddes, C. D. *J. Phys. Chem. B* **2005,** *109,* 3157.

(17) Zhao, B.; Cao, B.; Zhou, W.; Li, D.; Zhao, W. *J. Phys. Chem. C* **2010,** *114***,** 12517.